\documentclass[aps,prb,twocolumn,amsmath,amsfonts,superscriptaddress]{revtex4}
\usepackage[dvipdfmx]{graphicx}
\usepackage{dcolumn}
\usepackage{bm}
\usepackage{boxedminipage}
\usepackage{amsmath}
\usepackage{graphicx}
\usepackage{amsfonts}
\usepackage{feynmf}
\usepackage{mathrsfs}
\usepackage{braket}
\usepackage{listings}
\usepackage{algorithm}
\usepackage{algorithmic}
\usepackage{color}

\begin{document}


\title{Spectrum collapse of disordered Dirac Landau levels \\as topological non-Hermitian physics}


\author{Taiki Matsushita}
\affiliation{Department of Materials Engineering Science, Osaka University, Toyonaka, Osaka 560-8531, Japan}
\author{Yuki Nagai}
\affiliation{CCSE Japan Atomic Energy Agency, 178-4-4, Wakashiba, Kashiwa, Chiba, 277-0871, Japan}
\affiliation{Mathematical Science Team, RIKEN Center for Advanced Intelligence Project (AIP),
1-4-1 Nihonbashi, Chuo-ku, Tokyo 103-0027, Japan}
\author{Satoshi Fujimoto}
\affiliation{Department of Materials Engineering Science, Osaka University, Toyonaka, Osaka 560-8531, Japan}
\date{\today}

\begin{abstract}
We investigate disorder effects on Landau levels in Dirac electron systems with the use of a non-Hermitian quasiparticle Hamiltonian formalism. 
This formalism reveals that spin-dependent scattering rates induce the spectrum collapse of Landau levels, i.e., the disappearance of the energy gaps between $n$-th and $-n$-th levels under a finite external magnetic field. 
The spectrum collapse occurs in both weak and strong magnetic field regimes, thus showing a reentrant behavior.
Particularly, in the strong magnetic field regime, in contrast to naive expectation, 
the increase of a magnetic field stabilizes the spectrum collapse of Dirac Landau levels.
Furthermore, it is revealed that the spectrum collapse is associated with the emergence of a vortex texture with a topological winding number of a complex energy spectrum of the non-Hermitian system.
\end{abstract}


\maketitle


\section{Introduction}
Dirac electron systems are one of the central issues in condensed matter fields.~\cite{wehling_dirac}
Dirac electron systems attract many interests because of their unique properties.
Especially, the emergence of non-trivial topology in the Hilbert space has been extensively studied.~\cite{hasan_TI,sato_TSC}
The non-trivial topology manifests in topologically protected surface states and the associated novel transport phenomena.
For example, in the Haldane model, which is a well-known example of Dirac electron systems with non-trivial topology, the Hall conductivity is quantized as $\sigma_{xy}=ne^2/h\;\;(n\in \mathbb{Z})$ at low-temperature.~\cite{Klizing_QHE,TKNN_QHE,haldane_QHE}
The quantization of the transport coefficient originates from Berry flux and is characterized by the Chern number, which is a topological invariant.~\cite{kohmoto_QHE,XiaoBerry}
Another interesting example is graphene, which is a prototype of Dirac semimetals.
Graphene shows the unconventional quantization of the Hall conductivity under an applied magnetic field, $\sigma_{xy}=4e^2 (n+1/2)/h\;\;(n\in \mathbb{Z})$.~\cite{novoselov_QHEgraphene,zhang_QHEgraphene}
The unconventional quantization of the Hall conductivity originates from the "relativistic" energy spectrum and the valley degrees of freedom of Landau zero modes.
Dirac or Weyl points in Dirac bands can be regarded as sources or drains of Berry flux, and their existence leads to characteristic magnetotransport properties.~\cite{novoselov_QHEgraphene,zhang_QHEgraphene,Murakamiphase,Wantopological,yangAHE,HosurWSM}
For instance, Weyl semimetals have characteristic magnetotransport properties associated with the chiral anomaly, ~\cite{Nielsenanomaly}
such as the anomalous Hall effect, the chiral magnetic effect, and the negative magnetoresistivity.~\cite{FukushimaCME,GoswamiNMR,ZyuzinTQFT,GoswamiTQFT,Sonanomaly,li2016weyl}
The chiral anomaly in Weyl semimetals can be understood from the asymmetric energy dispersion of Landau zero modes. 
Disorder effects on Landau levels of Dirac bands are thus crucially important to understand these novel magnetotransport properties.~\cite{peres_electronic,zheng_disorder_hall,lee_disorder_WSM}

In general, the quasiparticle damping gives additional topological structures in materials related to non-Hermitian quantum systems with dissipation.~\cite{yoshida_NH2}
Quasiparticles in solid-state systems have damping effects due to potential scatterings. 
Generally, a disordered system can be regarded as a dissipative system with damping effects, which is described by a non-Hermitian Hamiltonian formalism.~\cite{kozii_NH}
This formalism was also used for understanding correlation effects in terms of topology of the quasiparticle complex energy eigenvalues of the non-Hermitian effective Hamiltonian.
Besides, the authors of Ref.~\onlinecite{kozii_NH} clarified that a spin (orbital)-dependent lifetime in Dirac electron systems leads to an exceptional band touching and the appearance of bulk Fermi arcs, which are characteristic features of non-Hermitian energy spectra.
This finding motivated various succeeding research works.~\cite{kozii_NH,nagai_HF,yoshida_NH,yoshida2019symmetry,TM_NH,michishita_NH,Zyuzin_NH,Zyuzin_NH2,kimura_NH,kimura_NH2,Okugawa_NH,papaj_NH,michishita_NH2,shen_QO}
In particular, Papaj and his coworkers demonstrated the validity of the quasiparticle Hamiltonian formalism for disordered systems by comparing the quasiparticle Hamiltonian approach to exact diagonalization analysis~.\cite{papaj_NH}
Y. Michishita {\it et al,.} addressed the origin of non-Hermitian effects in the quasiparticle Hamiltonian formalism, and revealed the relation between correlation effects in many body systems and dissipation effects in open quantum systems.~\cite{michishita_NH}
The quasiparticle Hamiltonian formalism is also useful for studying disorder effects on Landau levels.
This formalism is successful in revealing that the disorder-induced residual density of states possibly causes the quantum oscillation in Kondo insulators.~\cite{shen_QO}

As shown by these studies, the spin (orbital) dependent lifetime in Dirac electron systems leads to various intriguing physical properties.
In this paper, we discuss disorder effects on Landau levels in Dirac electron systems, which include Weyl semimetals and two-dimensional Dirac semimetals, with the use of the quasiparticle non-Hermitian Hamiltonian formalism.
It is found that the energy gap between Dirac Landau levels can disappear under a finite external magnetic field.
In particular, the stronger magnetic field stabilizes this effect more.
We refer to this counter-intuitive effect as the spectrum collapse of Dirac Landau levels.
The spectral collapse involves a vortex-like topological texture in the complex energy spectrum specific in non-Hermitian systems, which is characterized by a non-trivial winding number. 

In Dirac electron systems, spin-dependent scattering rates cause the spectrum collapse of Landau levels.
The collapse of Landau quantization occurs in both weak and strong magnetic field regimes, thus showing a reentrant behavior.
In the weak magnetic field regime, where the energy gaps between Dirac Landau levels are smaller than any other energy scales, 
it is rather difficult to distinguish between the spectral collapse characterizing non-Hermitian systems and the conventional broadening of spectra due to disorder.
However, in contrast,  in the strong magnetic field regime, where magnetic field dependent scatterings are dominant scattering processes, a stronger magnetic field stabilizes more the spectrum collapse of Landau levels. 
 Thus, the collapse of Dirac Landau levels shows a reentrant behavior.
Magnetic field dependent scattering rates, which originates from Landau zero-modes, are crucial for the spectrum collapse in the strong magnetic field regime.

This paper is organized as follows. 
In Sec.~\ref{formulation}, we show our model of Dirac electron systems.
Here, the quasiparticle Hamiltonian theory under an external magnetic field is presented.
In Sec.~\ref{Disordered Dirac LL}, we derive the expression of Landau levels and demonstrate the spectrum collapse of disordered Landau levels in Dirac electron systems.
In Sec.~\ref{Spectrum collapse}, we focus on the weak and strong magnetic field regimes to study the spectrum collapse, and demonstrates the reenrant behavior of the collapse of Landau levels.
Notably, we clarify that the Landau zero-modes perturbed by disorder cause the spectrum collapse in the strong magnetic field regime.
In Sec.~\ref{topo_transition}, we associate the spectrum collapse of Landau levels with the emergence of a topological winding number.


\section{Quasiparticle Hamiltonian formalism in disordered Dirac electron systems}
\label{formulation}
\subsection{Dirac Hamiltonian and Landau levels}
In this paper, we consider disorder effects on the following Dirac-type Hamiltonian
\begin{eqnarray}
\label{hamDSM}
H_0({\bm p})=\bm f(\bm p)\cdot \bm \sigma,
\end{eqnarray}
where $\bm \sigma=(\sigma_x,\sigma_y,\sigma_z)$ is the vector of Pauli matrices in spin space,
and $\bm f(\bm p)=(f_x(\bm p), f_y(\bm p), f_z(\bm p))$ is a vector function of momentum.
We focus on the vector function $\bm f(\bm p)$ with the following symmetry,
\begin{eqnarray}
f_x(\bm p)&=&-f_x(-\bm p),\\
f_y(\bm p)&=&-f_y(-\bm p),\\
f_z(\bm p)&=&f_z(-\bm p),
\end{eqnarray}
and assume a gapless condition $\min_{p_z} |f_z(p_z)|=0$. 
The Hamiltonian (\ref{hamDSM}) can describe various Dirac electron systems.
For instance, 
\begin{eqnarray}
\label{fvec_DSM}
\bm f(\bm p)=(\lambda p_x, \lambda p_y, 0),
\end{eqnarray} 
describes two-dimensional Dirac bands, which include surface states of topological insulators and graphene.~\cite{wehling_dirac,konig_TI,qi_TI}
On the other hand, 
\begin{eqnarray}
\label{fvec_WSM}
\bm f(\bm p)=(\lambda p_x,\lambda p_y, \gamma(p_z^2-m))\;(m>0),
\end{eqnarray} 
describes Weyl semimetals.~\cite{murakami_WSM,wan_WSM,okugawa_WSM,ozawa_WSM}
In this paper, we focus on these Dirac electron systems.

Let us introduce an external magnetic field along the $z$-direction ${\bm B} = (0,0,B_z)$ by assuming the minimal coupling ${\bm p}\to -i \nabla - {\bm A}\; (e=c=\hbar=1)$.
The Hamiltonian with the external magnetic field is given by
\begin{eqnarray}
\label{QP_Ham_DSM_wB}
H_0(-i \nabla  - {\bm A})=
\begin{pmatrix}
f_z(p_z)&&\lambda \pi_-\\
\lambda \pi_+&&-f_z(p_z)
\end{pmatrix},
\end{eqnarray}
where the covariant derivatives $\pi_i\equiv -i\partial_i-A_i \; (i=x,y),\; \pi_\pm \equiv \pi_x\pm i \pi_y$ are introduced. 
Here, we replaced $-i \partial/\partial z$ with the crystal momentum $p_z$ because the translational symmetry to $z$-direction is retained. 
The covariant derivatives satisfy $\left[ \pi_+,\pi_- \right]=2B_z$. 
Normalizing the covariant derivatives as $\hat{a}\equiv (2B_z)^{-1/2}\pi_+,\; \hat{a}^\dag \equiv (2B_z)^{-1/2}\pi_-$, 
we obtain the Bosonic creation and anihilation operators, which satisfy $\left[\hat{a},\hat{a}\right]=\left[\hat{a}^\dag,\hat{a}^\dag \right]=0,\; \left[\hat{a},\hat{a}^\dag \right]=1$.
The Bosonic operators recasts the Hamiltonian (\ref{QP_Ham_DSM_wB}) into
\begin{eqnarray}
\label{QP_Ham_DSM_wB1}
&&H_0(-i\nabla-{\bm A})=
\begin{pmatrix}
f_z(p_z)&&\sqrt{2B_z}\lambda \hat{a}^\dag\\
 \sqrt{2B_z}\lambda \hat{a}&&-f_z(p_z)
 \end{pmatrix}.
 \label{eq:effh_WSM_wB_arep}
 \end{eqnarray}
 We can easily diagonalize the Hamiltonian (\ref{QP_Ham_DSM_wB1}) with the eigenket $\ket{n} (n\geq0,n\in \mathbb{Z})$ of the number operator $\hat{n}=\hat{a}^\dag \hat{a}$.
The eigenvalues and the eigenfunctions with $|n|\geq 1$ are given by
\begin{eqnarray}
\label{eq: Gop}
H_0(-i\nabla-{\bm A})|\pm,n,p_z\rangle&=&E_{\pm,n}(p_z)|\pm,n,p_z\rangle,
\end{eqnarray}
\begin{eqnarray}
E_{\pm,n}(p_z)&=&\pm \sqrt{f_z(p_z)^2+2|B_z|\lambda^2n},\label{eq:epm} \\
|\pm,n,p_z\rangle&=&\begin{pmatrix}
\pm \frac{1}{\sqrt{2}}\left(1+\frac{f_z(p_z)}{E_{\pm,n}(p_z)}\right)^{\frac{1}{2}}|n\rangle\\
\frac{1}{\sqrt{2}}\left(1-\frac{f_z(p_z)}{E_{\pm,n}(p_z)}\right)^{\frac{1}{2}}|n-1\rangle
\end{pmatrix},
\end{eqnarray}
and the eigenvalues and the eigenfunctions with $n=0$ are given by
\begin{eqnarray}
\label{eq: Gop0}
H_0(-i\nabla-{\bm A})|0,p_z\rangle&=&f_z(p_z)|0,p_z\rangle,
\end{eqnarray}
\begin{eqnarray}
\label{eq: eigenket0}
|0,p_z\rangle&=&\begin{pmatrix}
|0\rangle\\
0
\end{pmatrix}.
\end{eqnarray}
At the momentum which $f_z(p_z)= 0$ satisfies, the Landau levels are described by $E_{\pm,n} \propto \sqrt{n}$, 
which is well known as the feature of Landau levels of Dirac bands.~\cite{wehling_dirac}

\subsection{Four dimensional description}

For the clarification of topological aspects of disordered Dirac Landau levels,
a four-dimensional description of the energy spectrum is quite useful.
We introduce a pseudo momentum $p_B \equiv \sqrt{2 |B_z|} \lambda$.
The energy gap between the $n$-th and $-n$-th Landau levels in the clean system is written as $\sqrt{n}p_B$, which describes the rigidity of Landau quantization.
With the pseudo momentum, the eigenvalues in Eq.~(\ref{eq:epm}) are expressed as, 
\begin{align}
 E_{\pm,n}(p_z,p_B) &= \pm \sqrt{f_z(p_z)^2+n p_B^2}.
\end{align}
We can regard the $(p_z,p_B)$-dependence of the Landau levels $ E_{\pm,n}(p_z,p_B)$ as the energy "dispersion" in the "momentum" space $(p_z,p_B)$. 
In this paper, we use this energy "dispersion" to discuss topological features of disordered Dirac Landau levels.

\subsection{Green's function and effective quasiparticle Hamiltonian}

Now, we introduce an effective quasiparticle non-Hermitian Hamiltonian to consider disorder effects. 
Disorder generates a self-energy $\Sigma(-i\nabla-{\bm A},\epsilon)$ in the impurity averaged Green's function.
The impurity averaged Green's function operator is defined as
\begin{align}
  {\cal G}(\epsilon) &=
  [\epsilon -H_0(-i\nabla-{\bm A}) - \Sigma(-i\nabla-{\bm A},\epsilon)]^{-1}.
\end{align}
The following effective quasiparticle Hamiltonian is useful to perform the expansion : 
\begin{align}
  \mathcal{H}_{\rm eff}(-i\nabla-{\bm A}) &\equiv H_0(-i\nabla-{\bm A}) + \Sigma(-i\nabla-{\bm A},0).
\end{align}
Here, we neglected the frequency dependence of the self-energy to focus on the low energy behavior.
The imaginary part of the self-energy makes this $2 \times 2$ "Hamiltonian" matrix non-Hermitian.
The eigenvalue problem of this non-Hermitian matrix $\mathcal{H}_{\rm eff}(p_z,n)$ is written as 
\begin{align}
    \mathcal{H}_{\rm eff}(-i\nabla-{\bm A}) |\sigma,p_z,n,;R \rangle &= \mathcal{E}_{\sigma,n}(p_z) |\sigma,p_z,n;R\rangle, \\
    \mathcal{H}_{\rm eff}^{\dagger}(-i\nabla-{\bm A})  |\sigma,p_z,n,;L \rangle &= \mathcal{E}^{\ast}_{\sigma,n}(p_z) |\sigma,p_z,n;L\rangle,
\end{align}
where $\langle \sigma,p_z,n;L| \sigma',p_z,m;R \rangle = \delta_{nm} \delta_{\sigma \sigma'}$.
The relation $\sum_{m,p_z} \sum_{\sigma}|\sigma,p_z,m;R \rangle \langle \sigma,p_z,m;L | = 1$ allows us to expand the Green's function operator as follow 
\begin{widetext}
\begin{align}
  {\cal G}(\epsilon) &= \sum_{m_1,m_2,p_{z1},p_{z2}} \sum_{\sigma_1,\sigma_2}
  |\sigma_1,p_{z1},m_1;R \rangle \langle \sigma_1,p_{z1},m_1;L|
  \frac{1}{\epsilon - \mathcal{E}_{\sigma_2,m_2}(p_{z2})}
  |\sigma_2,p_{z2},m_2;R \rangle \langle \sigma_2,p_{z2},m_2;L|,\\
  &= \sum_{n,p_{z}} \sum_{\sigma}
   |\sigma,p_{z},n;R \rangle G^{\rm eff}_{\sigma}(p_{z},n)
   \langle \sigma,p_{z},n;L|,
\end{align}
\end{widetext}
where the Green's function $G^{\rm eff}_{\sigma}(p_{z},n) \equiv \langle \sigma,p_{z},n;L|[\epsilon - \mathcal{H}_{\rm eff}(-i\nabla-{\bm A})]^{-1} |\sigma,p_{z},n;R \rangle$ is defined .
The complex eigenvalues of the quasiparticle Hamiltonian $\mathcal{H}_{\rm eff}(-i\nabla-{\bm A})$ are the poles of the Green's function. 
The Green's function, which the complex eigenvalues give, enables us to calculate any physical quantities, such as the spectral function measured by the ARPES.
The effective non-Hermitian Hamiltonian is thus useful to consider disorder effects on the ARPES spectra.

We can understand disorder effects with the quasiparticle Hamiltonian formalism.
This formalism revealed an exceptional band touching in disordered or correlated electron systems.~\cite{kozii_NH,yoshida_NH,papaj_NH,TM_NH,michishita_NH,yoshida_NH2,Zyuzin_NH,Zyuzin_NH2,michishita_NH2,kimura_NH,kimura_NH2,Okugawa_NH}
The exceptional band touching stems from non-Hermitian nature of the quasiparticle Hamiltonian.
The non-Hermiticity, which describes quasiparticle decay rates, makes it non-diagonalizable at a certain momentum.
This momentum is known as an exceptional point (EP) of eigenvalues in non-Hermitian systems.~\cite{shen_NHband,Gong_NH,kawabata_EPs}
The appearance of EPs leads to the exceptional band touching.
For instance, the appearance of EPs involves the bulk Fermi arc, the disorder-induced flat band, and the exceptional surface.
Thus, the non-Hermitian description is useful for understanding band structures of disordered or correlated electron systems with finite life-time of quasiparticles.


\section{Disordered Landau levels}
\label{Disordered Dirac LL}

\subsection{Impurity self-energy}
Let us discuss the impurity self-energy in the two-dimensional Dirac band systems with the $f$-vector (\ref{fvec_DSM}), and the Weyl semimetals with the $f$-vector (\ref{fvec_WSM}).
We assume magnetic impurities with the following short-range impurity potentials
\begin{eqnarray}
V_{\rm imp}(\bm x)=\sum_{\bm R_a}\left(V_{\rm nmag}\sigma_0+V_{\rm mag}\sigma_z\right)\delta(\bm x-\bm R_a),
\end{eqnarray}
where $\bm R_a$ is a spatial coordinate of an impurity, $V_{\rm nmag}$ and $V_{\rm mag}$ describe non-magnetic and magnetic impurity potentials, respectively.  
Here we set the spin polarization of impurities to the $z$-direction because the external magnetic field is assumed to be parallel to the $z$-axis, and the $x$ and $y$-spin components of the magnetic impurity potentials are suppressed.
Now, we assume a random impurity distribution.
With the impurity average, disorder effects are included in the impurity self-energy.

For Dirac electron systems with this impurity potential, the impurity self-energy has been well investigated so far.
We first determine the form of the impurity self-energy from the symmetry of the Hamiltonian~(\ref{hamDSM}), then discussing its physical origins.
The symmetry of the Hamiltonian restricts the form of the impurity self-energy.
In the case with the $\delta$-function type impurity potentials, the impurity self-energy is calculated from the local Green's function $G(\epsilon,\bm x, \bm x)$.
The momentum odd nature of the off-diagonal components $f_x$ and $f_y$ requires the diagonal form of the local Green's function, and thus the impurity self-energy becomes a diagonal matrix.~\cite{TM_NH}~\footnote{The higher-order scattering channels gives rise to the off-diagonal self-energy naturally.}
Finally, we obtain the following impurity self-energy
\begin{eqnarray}
\label{selfen}
\Sigma(\epsilon=0,p_B)=-i\Gamma_0(p_B) \sigma_0+i\Gamma_z(p_B) \sigma_z,
\label{eq:impself}
\end{eqnarray}
where $\Gamma_0(p_B),\; \Gamma_z(p_B)\in \mathbb{R}$.
Here, we neglected the real part of the self-energy because the real part just renormalizes chemical potential or changes the positions of Dirac (Weyl) points.
The self-energy~(\ref{selfen}) is independent of the crystal momentum because of the short-range character of the impurity potentials.
It is noted that the self-energy generally depends on $p_B$.
When $p_B=0$, the spin-dependent scattering rate, $i\Gamma_z (0) \sigma_z$-term, drastically changes the quasiparticle energy dispersion.
In two-dimensional Dirac electron systems, $i\Gamma_z (0) \sigma_z$-term splits a Dirac point into two EPs and generates the bulk Fermi arc, which connects two EPs.~\cite{kozii_NH}
In Weyl semimetals, $i\Gamma_z (0) \sigma_z$-term changes a Weyl point into an exceptional ring and a disorder-induced flat band.~\cite{Xu_WER}

Now, we discuss the physical origin of $i\Gamma_z\sigma_z$-term in these Dirac semimetals.
In the absence of a magnetic field, i.e. $p_B=0$, 
$i\Gamma_z (0) \sigma_z$-term stems from magnetic impurities in the case of two-dimensional Dirac electron systems, 
and also from multiple scattering processes due to non-magnetic impurities in the case of Weyl semimetals.~\cite{papaj_NH,TM_NH}
On the other hand, an external magnetic field drastically changes the impurity self-energy, and generates $i\Gamma_z (p_B) \sigma_z$-term 
generally in Dirac band systems, irrespective of whether impurities are magnetic or non-magnetic.
This is due to the asymmetric form of the wave function of Landau zero modes in the spin space (see Eq.~(\ref{eq: eigenket0})).
For example, in Weyl semimetals, the analysis based on the Born approximation with non-magnetic impurities ($V_{\rm mag}=0$) gives the following impurity self-energy~\cite{klier2015transversal} 
\begin{eqnarray}
\label{gamma0B}
\Gamma_0(p_B)&=&\frac{n_{\rm imp}V_{\rm nmag}^2 p_B^2}{8\pi v_F\lambda^2},\\
\label{gammazB}
\Gamma_z(p_B)&=&-\frac{n_{\rm imp}V_{\rm nmag}^2 p_B^2}{8\pi v_F\lambda^2}.
\end{eqnarray}
where $v_F=2\gamma \sqrt{m}$ is the Fermi velocity, and $n_{\rm imp}$ is the impurity density.
Notably, neither magnetic impurities nor multiple scatterings are unnecessary for the realization of the $i\Gamma_z (p_B) \sigma_z$-term in the case with a magnetic field.

\subsection{Quasiparticle complex eigenvalues with disorders}
\begin{figure}[b]
    \includegraphics[width=9cm]{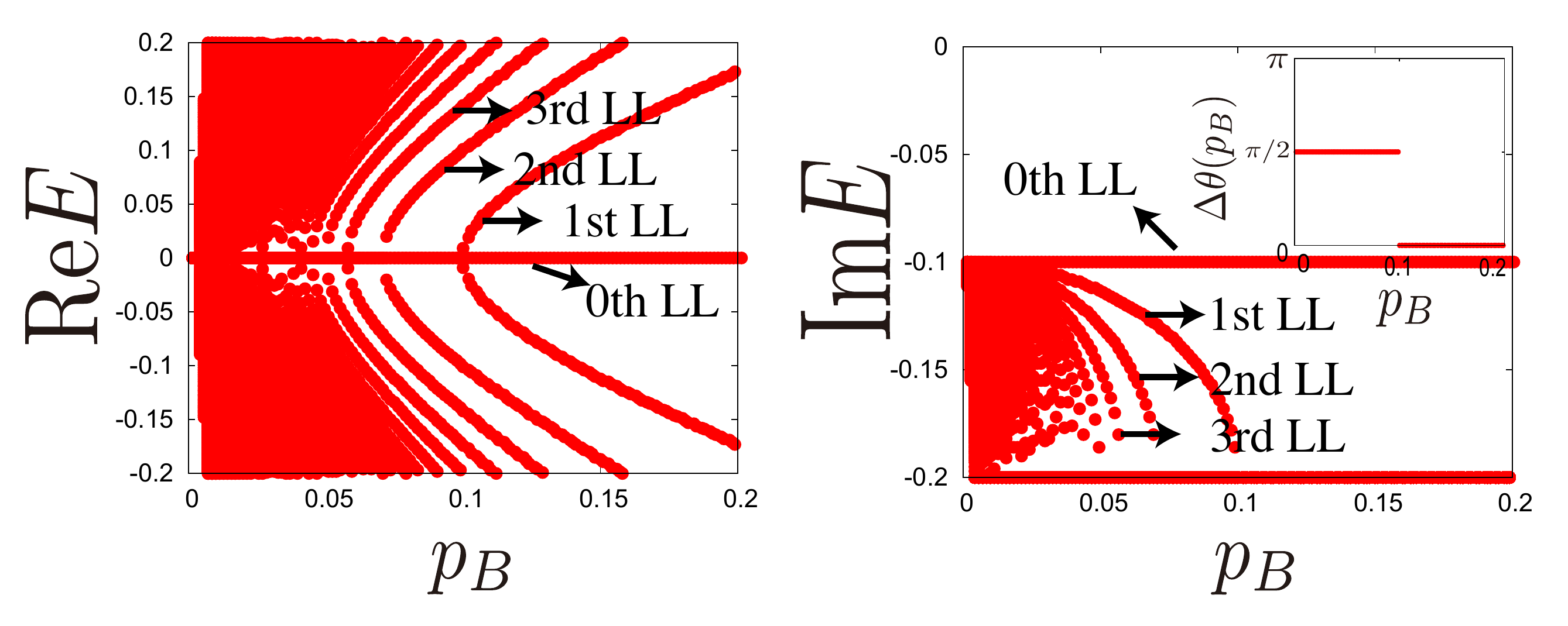}
    \caption{The complex energy dispersion in a disordered two-dimensional Dirac electron system with the magnetic field independent self-energy.
    The inset shows the vorticity $\Delta \theta_1(p_B)$. The parameters in all panels are : $\gamma_0=0.2,\; \gamma_z=0.1$.}
    \label{dispersion_Dirac}
\end{figure}

Next, we discuss disorder effects on Dirac Landau levels with the quasiparticle Hamiltonian formalism.
We assume the impurity self-energy (\ref{eq:impself}).
With the impurity self-energy, we obtain the following quasiparticle non-Hermitian Hamiltonian
\begin{widetext}
\begin{eqnarray}
&&\mathcal{H}_{\rm eff}(-i\nabla-{\bm A})=
\begin{pmatrix}
f_z(p_z)-i\Gamma_0(p_B)+i\Gamma_z(p_B)&&p_B \hat{a}^\dag\\
p_B \hat{a}&&-f_z(p_z)-i\Gamma_0(p_B)-i\Gamma_z(p_B)
 \end{pmatrix}.
 \label{eq:effh_WSM_wB_arep}
 \end{eqnarray}
\end{widetext}
This quasiparticle non-Hermitian Hamiltonian can easily be diagonalized.
The complex energy dispersions for the disordered Landau levels are given by
\begin{eqnarray}
\label{LL_Dirac_zeroth}
\mathcal{E}_{0}(p_z,p_B)&=&f_z(p_z)-i\Gamma_0(p_B)+i\Gamma_z(p_B),\\
\label{LL_Dirac}
\mathcal{E}_{\pm,n}(p_z,p_B)&=&-i\Gamma_0(p_B)\nonumber\\
&\pm& \sqrt{(f_z(p_z)+i\Gamma_z(p_B))^2+np_B^2},
\end{eqnarray}
where $n=1,2,3\cdots$.
\begin{figure*}[t]
    \includegraphics[width=16cm]{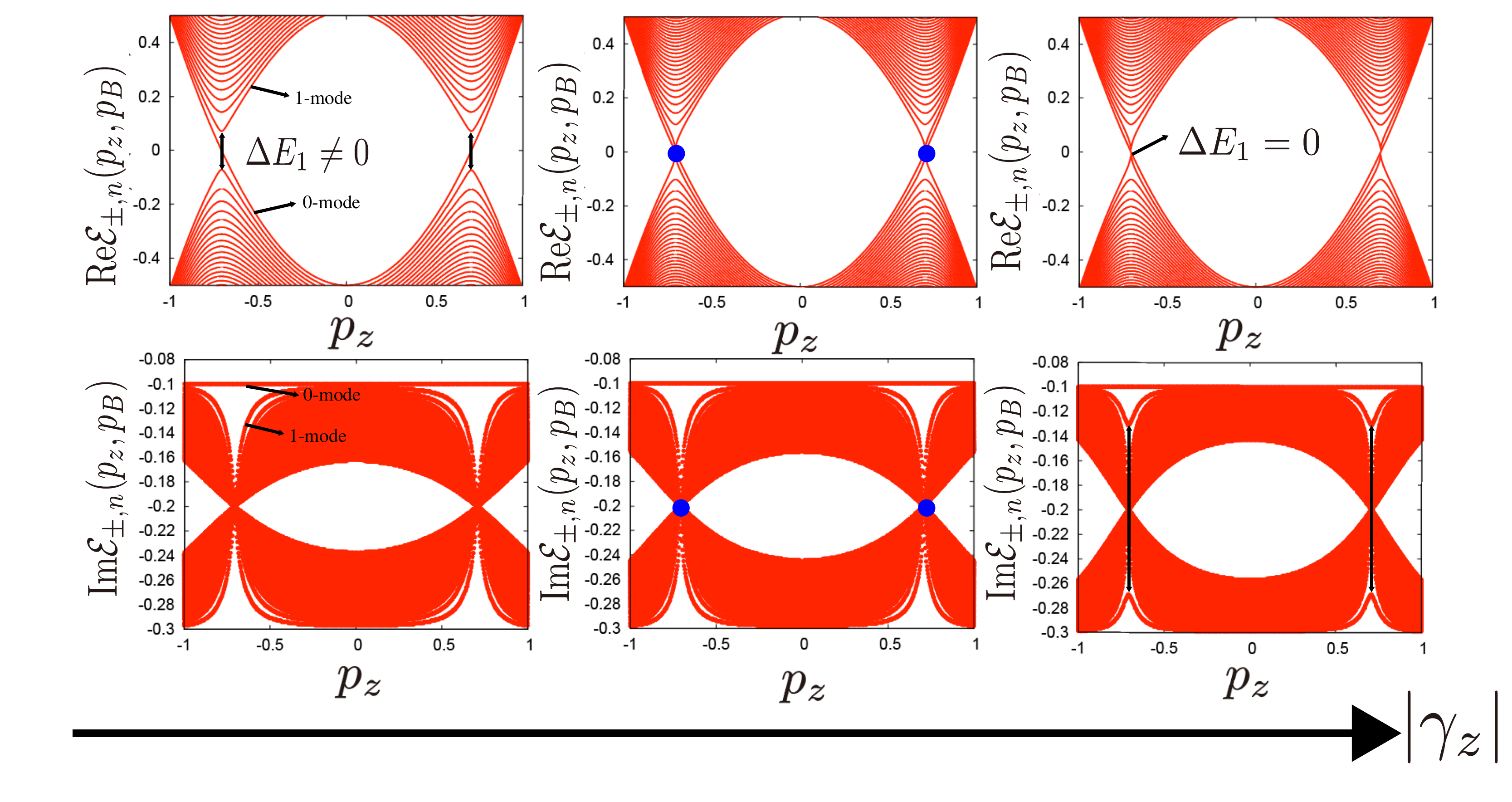}
    \caption{The complex energy dispersion in a disordered Weyl semimetal with the magnetic field independent self-energy.
The left and right panels are in the region where $\Delta E_1\neq 0$ and $\Delta E_1=0$, respectively.  The middle panels correspond to the emergence of the non-trivial winding number. }
\label{dispersion_Weyl_w/o_GammaB}
\end{figure*}
Eqs.~(\ref{LL_Dirac_zeroth}-\ref{LL_Dirac}) show that the $i\Gamma_z \sigma_z$-term changes the $n$-th $(|n|\geq 1)$ Dirac Landau levels drastically, whereas it just broadens the spectrum of the zeroth level.
To see this more precisely, we define the real energy gap between $n$-th and $-n$-th levels
\begin{eqnarray}
\label{real gap}
\Delta E_n &\equiv& \min_{p_z}{\rm Re}\left( \mathcal{E}_{+,n}(p_z,p_B)-\mathcal{E}_{-,n}(p_z,p_B) \right)\nonumber\\
&=&2{\rm Re}\sqrt{np_B^2-\Gamma_z(p_B)^2}\nonumber\\
&=&
\begin{cases}
2\sqrt{np_B^2-\Gamma_z^2(p_B)}\;\;\;\;\;{\rm for}\; np_B^2\geq \Gamma_z(p_B)^2,\\
0\;\;\;\;\;\;\;\;\;\;\;\;\;\;\;\;\;\;\;\;\;\;\;\;\;\;\;\;\;\;{\rm for}\; np_B^2<\Gamma_z(p_B)^2.
\end{cases}
\end{eqnarray}
Here, we used the gapless condition $\min_{p_z} |f_z(p_z)|=0$.
The real energy gap disappears when $|\Gamma_z(p_B)|$ is larger than a critical value $\sqrt{n}p_B$, which is the energy gap betweeen $n$-th and $-n$th levels in the clean system.
Thus, the real energy gap between Landau levels can disappear even in a nonzero magnetic field.
In FIGs.~\ref{dispersion_Weyl_w/o_GammaB} and~\ref{dispersion_Dirac}, the complex energy dispersion of a Weyl semimetal and a two-dimensional Dirac band system are shown. 
The calculated results shown in FIGs.~\ref{dispersion_Weyl_w/o_GammaB} and \ref{dispersion_Dirac} are obtained by assuming
the field-independent impurity self-energy $\Sigma(\epsilon=0)=-i\gamma_0 \sigma_0+i\gamma_z \sigma_z$ with $\gamma_0$ and $\gamma_z$ constants.
As seen in FIG.~\ref{dispersion_Weyl_w/o_GammaB}, $\Delta E_n$ decreases as $|\gamma_z|$ increases and disappears at a critical value of $|\gamma_z|$.
We refer to this disappearance of the real energy gap as {\it spectrum collapse of Landau levels}.
The field-independent self-energy is valid for weak magnetic fields.
However, we stress that the spectrum collapse of Landau levels occurs even for strong magnetic fields,
as clarified in the following section.

\section{Spectrum collapse of Dirac Landau level}
\label{Spectrum collapse}
In this section, we discuss the behavior of the collapsed Dirac Landau levels more precisely.
As Eq.~(\ref{real gap}) shows, the $i\Gamma_z\sigma_z$-term causes the collapse of Landau quantization.
Here, we assume the following forms of the impurity self-energy,
\begin{eqnarray}
\label{Gamma_0}
\Gamma_0(p_B)=\gamma_0+\gamma_{B0}p_B^2,\\
\label{Gamma_B}
\Gamma_z(p_B)=\gamma_z+\gamma_{Bz}p_B^2.
\end{eqnarray}
The magnetic field independent self-energy, $\gamma_0$ and $\gamma_z$, gives the impurity self-energy of the Dirac electron systems without magnetic fields.
The magnetic field dependent terms proportional to $\gamma_{B0}$ and $\gamma_{Bz}$, stems from the disordered Landau zero-modes.~\cite{klier2015transversal}

With this impurity self-energy, the real energy gap (\ref{real gap}) becomes,
\begin{eqnarray}
\label{real gap1}
\Delta E_n &=&2{\rm Re}\sqrt{np_B^2-\left(\gamma_z+\gamma_{Bz}p_B^2\right)^2}.
\end{eqnarray}
Here, we focus on the both weak and magnetic field regimes to understand more closely the features of the collapse of Dirac Landau levels.
 
We, first, consider the weak magnetic field regime.
In this regime, where $p_B$ is smaller than any other energy scale, the magnetic field dependent terms are negligible, which allows us to set 
$\Gamma_0(p_B)\simeq \gamma_0,\; \Gamma_z(p_B)\simeq \gamma_z$. 
Then, the real energy gap (\ref{real gap1}) is given by,
\begin{eqnarray}
\label{real gap weak pB}
\Delta E_n =
\begin{cases}
2\sqrt{np_B^2-\gamma_z^2}\;\;\;\;\;\;{\rm for}\; np_B^2\geq \gamma_z^2,\\
0\;\;\;\;\;\;\;\;\;\;\;\;\;\;\;\;\;\;\;\;\;\;\;\;{\rm for}\; np_B^2<\gamma_z^2.
\end{cases}
\end{eqnarray}
In the weak magnetic field regime, as the magnetic field increases, the energy gap becomes larger, suppressing
the spectrum collapse (see Fig.~\ref{dispersion_Dirac}).

\begin{figure}[b]
    \includegraphics[width=9cm]{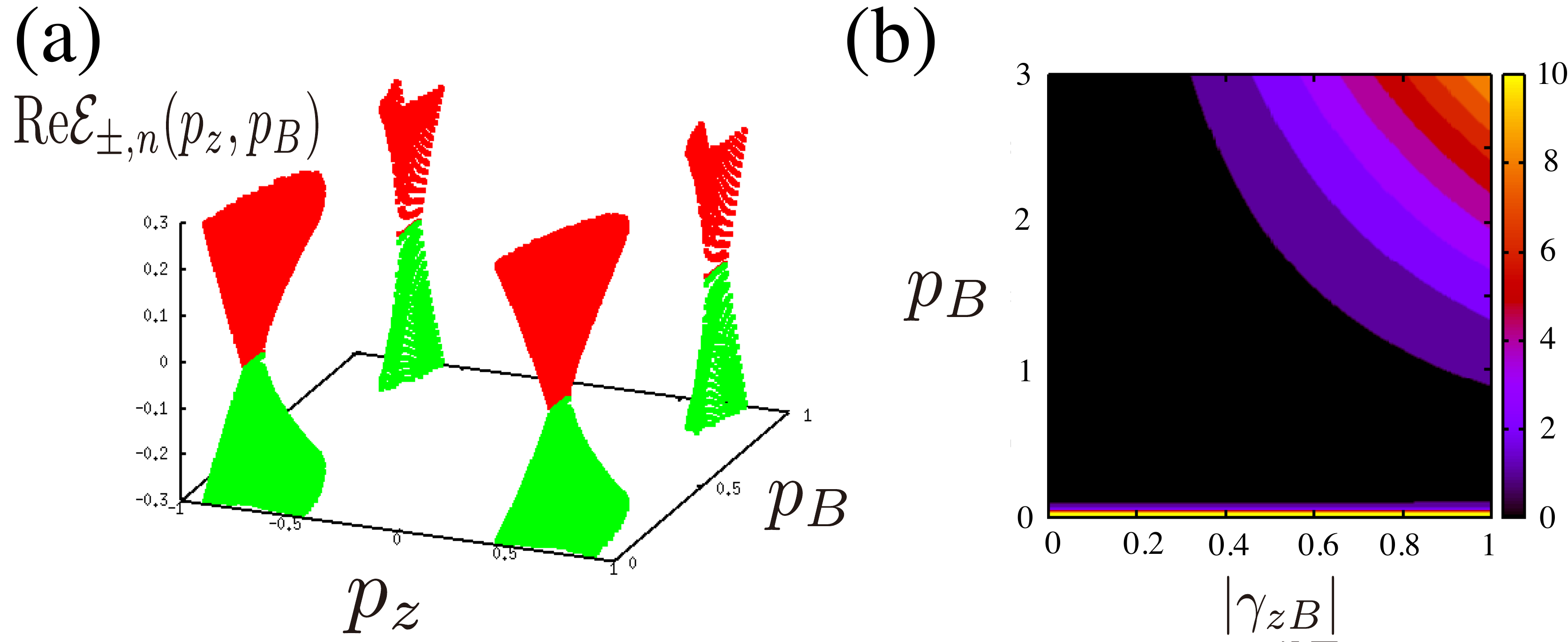}
    \caption{(a) The real part of the complex energy dispersion in the disordered Weyl semimetals with $|n|\geq1$.
    Here, we set $\gamma_0/\lambda=\gamma_z/\lambda=0.1,\;\gamma_{0B}=\gamma_{zB}=1.0$. (b) The number of the collapsed Landau levels with $\gamma_0/\lambda=\gamma_z/\lambda=0.1$.}
    \label{dispersion_Weyl w/GammaB}
\end{figure}

On the other hand, in the strong magnetic field regime, the behavior of the spectrum collapse is quite different.
In this regime, the magnetic field dependent term is dominant, and we can approximate the impurity self-energy as $\Gamma_0(p_B)\simeq \gamma_{B0}p_B^2,\; \Gamma_z(p_B)\simeq \gamma_{Bz}p_B^2$.
Then, the real energy gap (\ref{real gap1}) is given by
\begin{eqnarray}
\label{real gap strong pB}
\Delta E_n =
\begin{cases}
2p_B\sqrt{n-\gamma_{zB}^2p_B^2}\;\;\;\;\;\;{\rm for}\; n\geq \gamma_{zB}^2p_B^2,\\
0\;\;\;\;\;\;\;\;\;\;\;\;\;\;\;\;\;\;\;\;\;\;\;\;\;\;\;\;\;\;{\rm for}\; n<\gamma_{zB}^2p_B^2.
\end{cases}
\end{eqnarray}
In this regime, as the external magnetic field becomes larger, the spectrum collapse is more enhanced,
and the number of the Landau levels which collapse to the zero energy increases, in contrast with that in the weak magnetic field regime.
We can understand the spectrum collapse in the strong magnetic field regime by comparing the impurity self-energy with the energy gap in the clean case which is proportional to $p_B$.
As mentioned before, the magnetic field dependent self-energy originates from the disordered Landau zero modes and is proportional to the degeneracy of Landau levels $(\propto p_B^2)$.
It grows more rapidly than the gap in the clean systems for larger magnetic fields.
Thus, the collapse of the Dirac Landau levels exhibits a reentrant behavior as a function of an applied magnetic field (see Fig.~\ref{dispersion_Weyl w/GammaB}).
This behavior is a crucially different from trivially broaden Landau levels of disordered non-relativistic electron systems.

\section{Spectrum collapse as topological non-Hermitian physics}
\label{topo_transition}
\begin{figure}[t]
    \includegraphics[width=8cm]{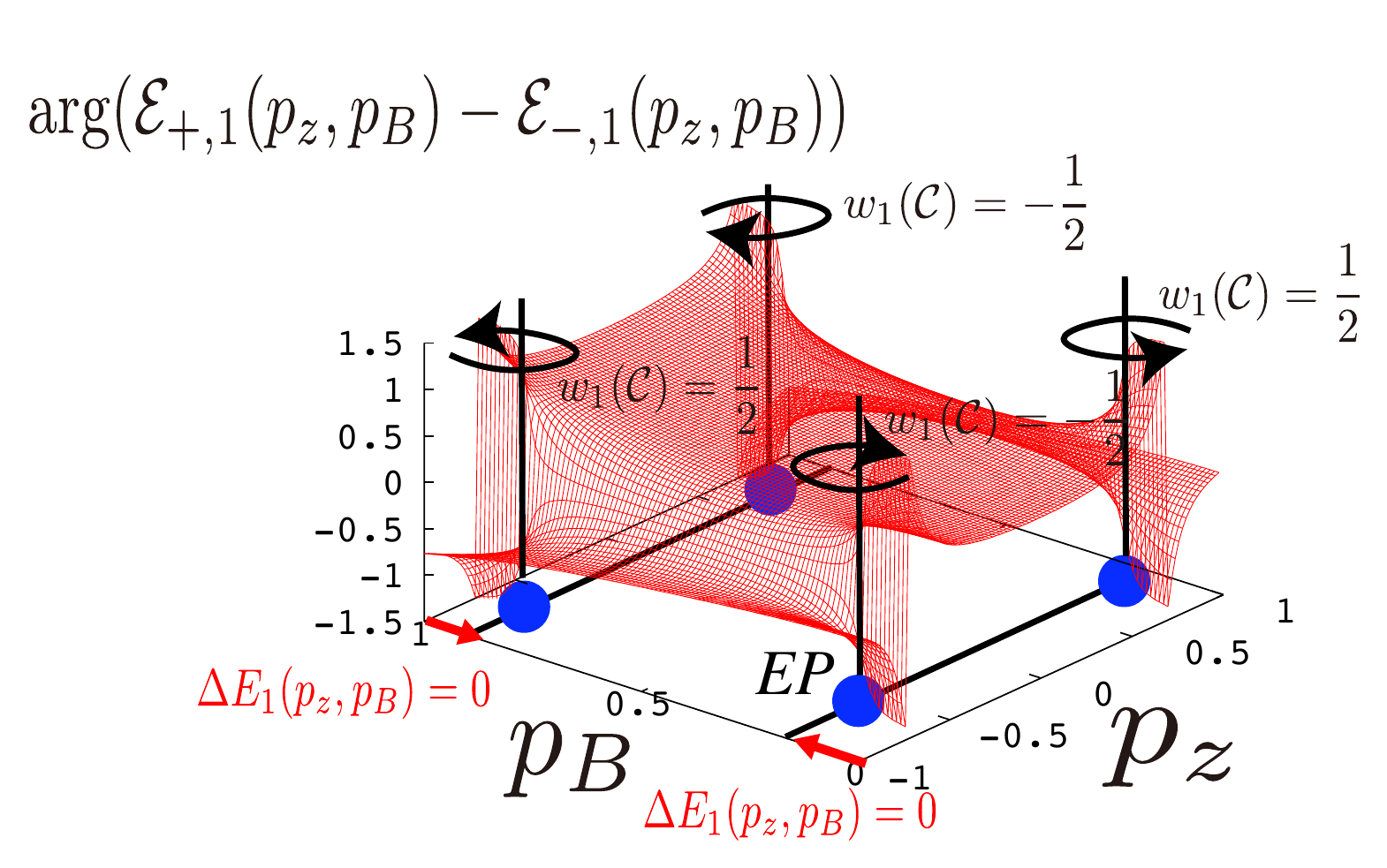}
    \caption{The phase difference of the complex energy between the 1-st and -1st Landau levels.}
    \label{phase_Weyl}
\end{figure}
Finally, we clarify the relation between the spectrum collapse of Dirac Landau levels and a topologically non-trivial winding number of the complex energy eignvalues.
The four-dimensional momentum approach, which is introduced in Sec.~\ref{formulation}, is useful to understand topological natures of the spectrum collapse.
First, let us consider the disordered Weyl semimetal with the $f$-vector (\ref{fvec_WSM}).
With the four-dimensional momentum description, the complex energy dispersion of disordered Landau levels in the Weyl semimetal is given by
\begin{eqnarray}
\label{LL_Weyl_pB}
\mathcal{E}^{\rm WSM}_{\pm,n}(p_z,p_B)&=&-i\Gamma_0(p_B)\nonumber\\
&\pm& \sqrt{(\gamma(p_z^2-m)+i\Gamma_z(p_B))^2+np_B^2}.\nonumber\\
\end{eqnarray}
Eq.~(\ref{LL_Weyl_pB}) shows that the gap closing points between the $n$-th and $-n$-th level ($\mathcal{E}_{+,n}(p_B)=\mathcal{E}_{-,n}(p_B)$) correspond to EPs in the momentum space with the synthetic dimension.
EPs in the momentum space involve the non-trivial winding number, which is defined as
\begin{eqnarray}
\label{Winding_Num}
  w_n(\mathcal{C})&=&\frac{1}{2\pi}\oint_\mathcal{C} \frac{\partial}{\partial \bm p} {\rm arg}(\mathcal{E}_{+,n}(p_z,p_B)-\mathcal{E}_{-,n}(p_z,p_B))\cdot d\bm p.\nonumber\\
\end{eqnarray}
Here, the differential operator is defined as $\frac{\partial}{\partial \bm p}=(\frac{\partial}{\partial p_z},\frac{\partial}{\partial p_B})$, and the closed path $\mathcal{C}$ is in the two-dimensional momentum space $(p_z,p_B)$.
As seen in FIG.~\ref{phase_Weyl}, the spectrum collapse involves the emergence of the winding number $w_n(\mathcal{C})=\pm \frac{1}{2}$ in the momentum space.

This argument is also applicable to two-dimensional Dirac electron systems.
In the two-dimensional Dirac electron system with the $f$-vector (\ref{fvec_DSM}), the complex dispersion of disordered Landau levels is given by,
\begin{eqnarray}
\label{LL_Dirac_pB}
\mathcal{E}^{\rm DSM}_{\pm,n}(p_B)&=&-i\Gamma_0(p_B) \pm \sqrt{-\Gamma_z(p_B)^2+np_B^2}.
\end{eqnarray}
As in the case of Weyl semimetals, the gap closing points ($\mathcal{E}_{+,n}(p_B)=\mathcal{E}_{-,n}(p_B)$) correspond to EPs in the one-dimensional momentum space.
The phase difference between $n$ and $-n$-th level,
\begin{eqnarray}
 \Delta \theta_n(p_B)=\arg (\mathcal{E}_{+,n}(p_B)-\mathcal{E}_{-,n}(p_B)),
\end{eqnarray}
is useful for understanding topological natures of the collapse of Dirac Landau levels.
As shown in the inset of FIG.~\ref{dispersion_Dirac}, the phase difference $\Delta \theta_1(p_B)$ involves a discontinuous behavior at the gap closing points.
At this point, $\Delta \theta_1(p_B)$ changes by $-\pi/2$, and thus EPs in one-dimensional momentum space correspond to half-vortices.

The half-vortices are also related to the winding number (\ref{Winding_Num}), as describe below.
We can regard the one-dimensional dispersion (\ref{LL_Dirac_pB}) as a projection from the two-dimensional dispersion (\ref{LL_Weyl_pB})
to the one-dimensional space with $p_z^2=m$.
From this perspective, the half-vortices can be understood as the projection of the winding of the phase of the complex quasiparticle energy in the two-dimensional momentum space.
The topological winding number (\ref{Winding_Num}) guarantees the robustness of the phase difference in the two-dimensional Dirac band systems.
Thus, this discontinuous phase difference does not vanish unless a half vortex and an anti-half vortex annihilate pairwisely.

\section{Conclusion}
This paper addressed disorder effects on Landau levels in Dirac band systems with the non-Hermitian quasiparticle Hamiltonian formalism.
With this formalism, the spectrum collapse of Landau levels was established.

We focused on the weak and strong magnetic field regimes to clarify precisely the characters of the collapse of Dirac Landau levels.
It is found that a reentrant behavior of the spectrum collapse of Dirac Landau levels appears.

Moreover, we elucidate that the spectrum collapse is intimately related to the topological winding number (\ref{Winding_Num}) 
of vortex texture in a complex energy eigenvalue, which is a characteristic feature of non-Hermitian systems.

Finally, we discuss how to detect the spectrum collapse.
The lattice strain in Dirac electron systems gives rise to an elastic gauge field, which quantizes the energy dispersion in the same manner as the magnetic field.~\cite{guinea_strain,pikulin_strain}
The recent ARPES study succeeded in measuring Dirac Landau levels in strained Dirac electron systems.~\cite{levy_strain,nigge_strain}
In two-dimensional Dirac electron systems, such as graphene, the strain-induced fictitious magnetic field naturally appears through the interaction with  the substrate.
Thus, an ARPES measurement is a promising way to study the collapse of Landau quantization.
However, a difficulty is how to distinguish the collapse of Dirac Landau levels from trivial broadening of spectra due to disorder.
We would like to stress that the reentrant behavior of the collapse of Dirac Landau levels found in this study is a remarkable signature of
the spectrum collapse of Landau levels arising from non-Hermitian physics, which allows us to differentiate 
between the topological non-Hermitian effect and trivial broadening.
The experimental verification of this effect is an interesting future issue.

 \section*{Acknowlegement} 
T.M.\ thanks K. Nomura and Y. Michishita for invaluable discussion. 
T.M.\ was supported by a JSPS Fellowship for Young Scientists. 
This work was partly supported by the Grant-in-Aids for Scientific
Research from MEXT of Japan [Grants No. 17K05517, and KAKENHI on Innovative Areas ``Topological Materials Science'' [No.~JP15H05852]  and "J-Physics" [No.~JP18H04318], and JST CREST Grant Number JPMJCR19T5, Japan.

\begin{appendix}

\section{$i\Gamma_z \sigma_z$-term in Elastic gauge field}
In this paper, we consider effects of the impurity self-energy on Dirac Landau levels, and mainly focus on the the spin-dependent scattering rates $i\Gamma(p_B)\sigma_z=i(\gamma_z+\gamma_{Bz}p_B^2)\sigma_z$. 
The $i\gamma_z \sigma_z$-term arises from magnetic field independent scattering processes due to magnetic impurities or multiple scatterings.
On the other hand, the $i\gamma_{zB}p_B^2\sigma_z$-term is due to magnetic field dependent scattering processes.

It is noted that
an elastic gauge field also gives rise to the $i\gamma_{zB}p_B^2\sigma_z$-term in the same manner as an applied magnetic field.~\cite{klier2015transversal}
To see this, let us consider the following linear Hamiltonian
\begin{eqnarray}
H_{\tau_z}=\tau_z v_F(-i\nabla)\cdot {\bm \sigma},
\end{eqnarray}
where $\tau_z$ is a chirality index.
In the case with the elastic gauge field ${\bm A}^5$, the Hamiltonian of a Dirac band system becomes,~\cite{guinea_strain,pikulin_strain}
\begin{eqnarray}
H_{\tau_z}=\tau_z v_F(-i\nabla-\tau_z{\bm A}^5)\cdot {\bm \sigma}. 
\end{eqnarray}
The Hamiltonian with the elastic magnetic field is the same as the Hamiltonian with a magnetic field except for the coupling charge $\tau_z$.
Thus, we can straightforwardly derive the $i\gamma_{zB}p_B^2\sigma_z$-term of the self-energy for the elastic gauge field in the same manner as that for a magnetic field.
\end{appendix}

\
\bibliography{referencenonhermiteWeyl}
\bibliographystyle{apsrev}
\end{document}